\documentclass{aastex63}

\usepackage{amsmath}
\published{Astronomy Reports, Volume 62, Issue 10, pp.654-663, 2018, DOI:  10.1134/S1063772918100037}

\shorttitle{Helical jet of the blazar S5 0716+714}
\shortauthors{Butuzova M.S.}

\begin{document}

\title{Periods of the Long-Term Variability of the Blazar 0716+714
and \\Their Inter-Correlations in a Helical Jet Model}

\correspondingauthor{Marina Butuzova}
\email{mbutuzova@craocrimea.ru}

\author[0000-0001-7307-2193]{Marina S. Butuzova}
\affiliation{Crimean Astrophysical Observatory of RAS, 298409, Nauchny, Russia}

\begin{abstract}
Various quasi-periods for the long-term variability of the radio emission, optical emission, and
structural position angle of the inner part of the parsec-scale jet in the blazar 0716+714 have been detected.
The relationships between these quasi-periods are interpreted assuming that the variability arises due to
helical structure of the jet, which is preserved from regions near the jet base to at least 1 milliarcsecond from
the core observed in radio interferometric observations. The radiating jet components should display radial
motions with Lorentz factors of $\approx3$, and decelerate with distance from the jet base. The best agreement with
the data is given in the case of non-radial motions of these components with a constant physical speed. It
is also shown that the helical shape of the jet strongly influences correlations both between fluxes observed
in different spectral ranges and between the flux and position angle of the inner part of the parsec-scale jet.
\end{abstract}


\section{Introduction} 
\label{sec:intro}
Blazars are a class of Active Galactic Nuclei
(AGN) whose relativistic parsec-scale jets are oriented
close to the line of sight. Therefore, the flux
density emitted by the jet is enhanced by relativistic
effects in the observer’s frame, so that it dominates
the emission of other parts of the AGN. This may
be able to explain the fact that no lines are observed
in the spectrum of the blazar 0716+714. Indirect
estimates of its redshift have yielded $z\approx 0.31$ \citep{Bychkova06,Nilsson08}, $0.2315 < z < 0.3407$ \citep{Danforth13}, and $z\geq 0.52$ \citep{Sbarufatti05}.
This object is variable over the entire electromagnetic
spectrum, on both short and long time scales \citep[see,
e.g.,][]{Wagner96,Raiteri03,WuZhou07,Poon09,Gorshkov11,Volvach12,Rani13,Liao14,Bychkova15}.

0716+714 has also been observed with Very Long
Baseline Interferometry (VLBI) in the framework of
a 2~cm survey using the Very Long Baseline Array
(VLBA), and monitored as part of the MOJAVE
project \citep[see, e.g.,][]{Bach05,Lister13,Pushkarev17}.
The bright compact
VLBI core visible in these maps is the region of
the jet where the medium becomes optical thin to
radiation at the given wavelength. For more than
150 sources (including 0716+714), the position of the
VLBI core has been found to shift closer to the jet
base with increasing frequency \citep{Pushkarev12}, interpreted as an effect of synchrotron self-absorption in the jet \citep{MarcaideShap1984,Lobanov98,KovalevLobPu08}.
That is, the magnetic-field strength and density
of the radiating particles decrease with distance from
the jet base, so that the medium becomes optically thin to radiation at increasingly lower frequencies.
Synchrotron self-absorption could also be responsible
for the observed delays between flares observed at
different radio frequencies on single dishes \citep[see, e.g.,][]{KudryavtsevaGab11,AgarwalMohan17}.
Without interpretation by synchrotron selfabsorption
a number of authors have found time delays
between variability at different radio frequencies
and in different spectral ranges.

For example, \citet{Raiteri03} found that the delay
between the variability of 0716+714 at 22$-$23~GHz
and 15~GHz is 6$-$9 days, between 22$-$23 and 8~GHz
is 22$\pm$2$^\text{d}$, and between 15 and 5~GHz is 53$\pm$2$^\text{d}$.
No reliable correlation between the optical and 15~GHz
fluxes has been found. However, \citet{Rani13} found
a correlation between the V and 230~GHz fluxes with
a time delay of $\approx65^\text{d}$.
The delay between the optical
and gamma-ray ranges is $\approx1.4^\text{d}$ \citep{Larionov13}.
The millimeter wave length
variability lags the gamma-ray variability
by $82\pm32^\text{d}$ \citep{Rani14}.
\citet{WuZhou07} and \citet{Poon09} attempted to determine the delays between the variability of 0716+714 in different optical bands. If there
are such delays, they are less than the time resolution
of those observations.

Thus, the flux variability at low frequencies is delayed
compared to the variability at higher frequencies.
Analyses of long-term series of observations
reveal different quasi-periods\footnote{Here and below, quasi-periods refer to variability periods detected in certain time intervals at a specified confidence
level using specialized methods such as discrete correlation
functions \citep{EdelsonKrolik88}, structure functions \citep{Simonetti85}, and others.} for the long-term optical ($\approx3.3$~yrs \citep{Raiteri03}) and radio (5.5$-$6~yrs \citep{Raiteri03,Bychkova15,LiuMi12} variability, and also for variations in the position angle of the inner jet PA$_\text{in}$ ($\approx11$~yrs \cite{Lister13}).

The variations of PA$_\text{in}$ can be explained in a natural
way if the jet has a helical shape \citep{Bach05,Lister13}, which also
often provides the simplest interpretation of a variety
of observed properties of AGNs.
For example, in the
case of 0716+714, variations in the spectral energy
distribution \citep{OstoreroRait01} and the kinematics of features in the
parsec-scale jet \citep{Butuzova18} have both been explained in this
way. The helical jet shape also gives rise to periodic
variations in the viewing angle of the radiating
regions, leading to corresponding variations of the
Doppler factor, which should be manifest as longterm
periodicity of the flux variations. The differences
in the long-term periods for the radio variability, optical
variability, and PA$_\text{in}$ variations can be explained
by either an overall deceleration of the speed of the
jet or non-radial motions of the radiating regions in
the jet. The latter is most probable for 0716+714
(see Section~2). Section~3 presents interpretations
of the following results. The first is the opposite
results obtained in searches for correlations between
the radio and optical flux variations using the same
statistical method \citep{Raiteri03, Rani13} The second is the fact
that time intervals when there is a strong positive
correlation between PA$_\text{in}$ and the gamma-ray flux
alternate with intervals in which there is a strong
negative correlation between these quantities \citep{Rani14}. A
discussion of the obtained results and our conclusions
are presented in Section~4.

\section{Relationship between the long-term variability periods for various observed quantities}

Analysis of many-year radio and optical light
curves of the blazar 0716+714 indicate the presence
of quasi-periods in the long-term variability. Optical
data for 1994$-$2001 indicated a period of about
3.3~yrs \citep{Raiteri03}. This period was never confirmed, possibly
because other studies considered data obtained over
shorter time intervals. For example, \citet{Rani13} did not detect any reliable periods for the long-term
variability in their analysis of data for 2007$-$2010.
It is difficult to detect long-term periodicity in the
optical due to the superposition of short-term flares
on the longer-term trend. In the radio, where the flare
component is weaker, variability periods of 5.6$-$6~yrs
at 14.5 and 15~GHz for 1978$-$2001 \citep{Raiteri03}, 5.5$-$6 yrs at
22~GHz for 1992.7$-$2001.2 \citep{Bach05}, and $5.8 \pm 0.4$~yrs at
15~GHz for data after 2001 \citep{LiuMi12} have been found.

Data from more than 30 years of observations carried
out on telescopes of the Crimean Astrophysical
Observatory (CrAO), the Metsahovi Radio Observatory,
and the University of Michigan Radio Astronomy
Observatory at frequencies from 4.8 to 36.8~GHz 
also indicate the presence of a period of $\approx8$~yrs,
together with shorter periods \citep{Bychkova15}.
Quasi-periodicity is also present in the variations of the position angle
of the inner parsec-scale jet of 0716+714. Analysis
of data for 26 epochs of observations (from 1992.7)
at 2.9, 8.4, 15.3, and 22.2~GHz led to the detection
of variations in the position angle of the jet lying
within 1~mas from the VLBI core
with a period of $7.4\pm1.5$~ yrs and an amplitude of
$3.5^\circ$ \citep{Bach05}. 
Observations at 15~GHz obtained from
1994.5$-$2011.5 displayed a period for the PA$_\text{in}$ variations
of 10.9~yrs and an amplitude of $11^\circ$ (in this
case, for the mean position angle of all jet features
at distances $0.15-1$~mas from the core, weighted
according to their flux densities) \citep{Lister13}.

Therefore, the lack of agreement between the periods
for the radio and optical variability seems to
suggest that the brightness variations of the blazar
0716+714 in these spectral ranges are associated
with physical processes occurring in its jet. On the
other hand, the periodic variations of the position
angle of the inner jet testify to periodic variations in
the jet direction, leading to variations in the spectral
flux density, since
\begin{equation}
    F_\nu\propto \nu^{-\alpha} \delta^{s+\alpha},
    \label{eq:eq1}
\end{equation}
where $\nu$ is the observing frequency, $\alpha$ the spectral
index, $s = 3$ if the depth of the radiating region can be
neglected ($s = 2$ otherwise), and the Doppler factor is
\begin{equation}
    \delta=\delta\left(\theta, \beta \right)=\sqrt{1-\beta^2} \left(1-\beta \cos \theta \right)^{-1}.
    \label{eq:eq2}
\end{equation}
Here, $\theta$ is the angle between the velocity vector for a
jet component and the line of sight at the given time,
and $\beta$ is the physical speed of the radiating feature in
units of the speed of light $c$. If $\beta$ is constant, periodic
variations of $\theta$ will give rise to long-term variability of
the radio and optical flux densities and of PA$_\text{in}$ with
the same period. However, the periods for these three
types of variations are different. We will elucidate
the origins of this contradiction under the hypotehsis
that the jet is helical in shape. We will explore this
using the schematic representation of a jet comprised
of individual components forming a helical line on the
surface of an notional cone \citep{Butuzova18}.
This corresponds to results of recent studies based
on stacked VLBI images for individual sources \citep{Pushkarev17},
which indicate that, for many sources, including
0716+714, the jet features on scales from hundreds
to thousands of parsecs are located inside a cone. We
take jet components to be individual radiating regions
of the jet that become observable when they reach
distances from the VLBI core of $\lesssim0.1$~mas. For our
subsequent arguments, it is not important whether
these components are regions of enhanced particle
density or shocks where electrons are accelerated and
subsequently injected into the surrounding space.
The position of a component on the surface of this
cone can be described by an azimuth angle $\varphi$
measured along a circular arc formed by a planar
cross section of the cone perpendicular to its axis.
(A detailed schematic of the jet and its geometrical
parameters are described by \citet{Butuzova18}).
The coordinate origin for $\varphi$ was taken to be the point located in the
plane of the line of sight and the cone axis, on the far
side of the cone relative to the observer.

\begin{figure}
    \centering
    \includegraphics[scale=0.8]{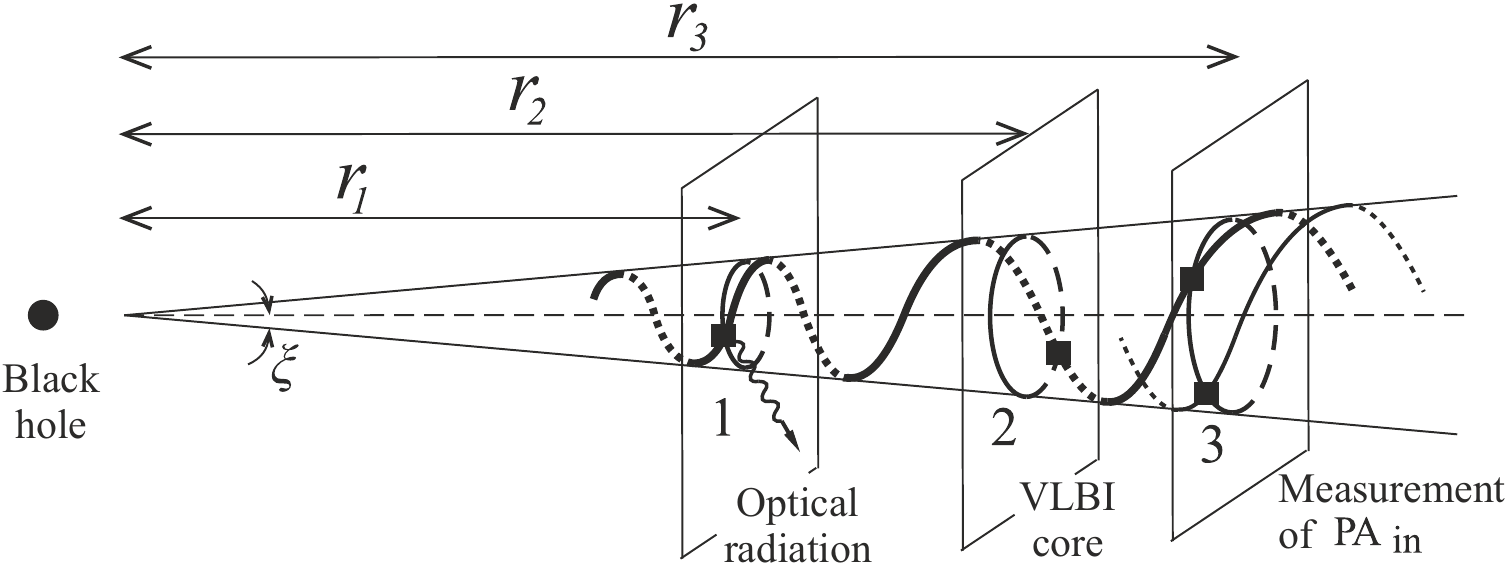}
    \caption{Schematic of the arrangement of regions in the jet in which the observed optical and radio emission arise and the region
in which PA$_\text{in}$ becomes measureable (the regions where a jet component crosses the circles \textit{1}, \textit{2}, and \textit{3}, respectively). The helical jet (without details of its components) is presented at a specified time by the thick bold curve. The thin curve shows the position of part of the jet in the vicinity of circle \textit{3} at some subsequent time. We can see that the point where the jet crosses circle \textit{3} shifts with time (similarly for circles \textit{1} and \textit{2}). Over the variability periods for the optical emission, radio emission or
PA$_\text{in}$, these points undergo a full revolution about circles \textit{1}, \textit{2}, and \textit{3}, respectively.}
    \label{fig:fig1}
\end{figure}

Taking into account synchrotron self-absorption,
we assumed that the medium becomes transparent
to the optical radiation of a jet component when it
reaches circle~\textit{1} (Fig.~\ref{fig:fig1}) formed by the cross section
of the cone by a plane orthogonal to the cone axis at a
distance $r_1$ from its apex. Continuing from the active
nucleus, the jet component reaches the analogously
formed circle~\textit{2} at the distance $r_2$.
In this region, the medium becomes transparent to the radio emission
of the jet formed in its VLBI core (at the given
frequency). Moving farther, the component reaches
circle~\textit{3} at a distance $r_3$ from the cone apex, where
it is manifest on VLBI maps as the closest component
to the core. We took this to be the distance at
which PA$_\text{in}$ is measured. For each of these circles,
we introduced a notional point moving such that
it coincides with the position of the jet component
intersecting the corresponding circle at a given time.
As follows from formulas (\ref{eq:eq1}), (\ref{eq:eq2}), the periods of the
optical and radio variability will then be equal to the
period of rotation of notional points around circles~\textit{1}
and \textit{2}, respectively.
Figure~\ref{fig:fig1} shows that the period for
variations of PA$_\text{in}$ will similarly be equal to the period
of rotation of a notional point around circle~\textit{3} \citep[see
also][Fig.~2]{Butuzova18}. In order for the helical structure of
the jet to be preserved over a long time, we assumed
that the speed of the components was constant, or
at least that this speed varies with distance from the
active nucleus in the same way for all components.

\subsection{Radial Motion of the Jet Components}

\begin{figure}
    \centering
    \includegraphics{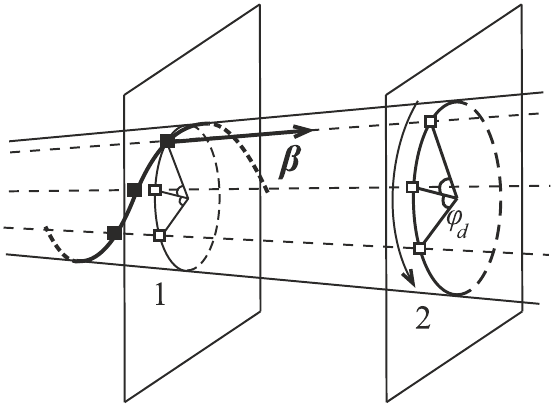}
    \caption{Schematic for interpreting the variability periods in the case of ballistic motion of the jet features. Part of the helical jet in the vicinity of the circle \textit{1} is shown by the bold curve. The filled squares represent several jet components. The trajectories of each of the components lie along the generator of the cone (dashed curves). The hollow squares show the positions where these components will intersect the circles \textit{1} and \textit{2}. The arrow indicates the direction of motion of a notional point along the circle \textit{2} (similarly for the circle \textit{1}).}
    \label{fig:fig2}
\end{figure}

We will first consider the case when the jet components
move outward along the generating cone (so-called
radial, or ballistic motion, see Fig.~\ref{fig:fig2}).
Without loss of generality, we can assume that the difference
in the azimuth angles of each of two successive components
is some value $\varphi_d$ in radians.
We denote $\Delta t^\prime_1$ to be the time interval in the comoving frame between
the times when any two successive components cross circle~\textit{1}.
Over this time, a notional point moving
along circle~\textit{1} describes an arc with length $\varphi_d$. 
For simplicity, we assumed that $2 \pi$ is a multiple of $\varphi_d$.
In this case, some number $n$ of components cross
circle~\textit{1} over the rotation period of the notional point.
Since the interval between two events in the
observer's frame is smaller than the interval in the
source rest frame by a factor $\delta$,
\begin{equation}
    \Delta t^\prime=\delta \Delta t,
    \label{eq:eq3}
\end{equation}
the variability period for the optical emission in the
observer's frame will be
\begin{equation}
    P_1=\sum_{j=1}^{n} \frac{\Delta t'_1}{\delta \left( \theta \left( \varphi\right)\right)} \approx \frac{n \Delta t'_1}{\delta \left(\theta_0, \beta \right)}.
    \label{eq:eq4}
\end{equation}
The right-hand side of (\ref{eq:eq4}) was obtained as follows.
The angle $\theta$ of the components crossing circle~\textit{1} over
the period varies from $\theta_0-\xi$ to $\theta_0+\xi$, where $\theta_0$ is
the angle between the cone axis and the line of sight.
According to formula (\ref{eq:eq2}), the Doppler factor varies
cyclically in some interval. Deviations of the Doppler
factor from its mean value can be neglected, since
their magnitudes are not large, due to the smallness
of the angle $\xi$, and their sum over the period is zero.
For our further estimates, we took the mean Doppler
factor to be $\delta\left(\theta_0, \beta \right)$.

Moving farther, the component crosses circle~\textit{2}.
The variability period for the radio emission $P_2$ can
be written similarly to (\ref{eq:eq4}), but with the subscript ``1''
replaced with ``2''. Due to the character of the motion
(Fig.~\ref{fig:fig2}), the azimuthal angle of each jet component
does not vary with time. The number of components
crossing circles~\textit{1} and \textit{2} during the variation period
$\varphi$ is also constant. 
The time intervals between the
moments of intersection by two successive components
of circles \textit{1} and \textit{2} are equal $\left(\Delta t^\prime_1=\Delta t^\prime_2 \right)$.
In the absence of deceleration of the components, we
should have $P_2=P_1$. Since this is not observed, we
supposed that the speed of the components crossing
circle~\textit{2} was $\beta_2=a \beta_1$. Here, $0 < a < 1$ and $a > \beta_2$
(otherwise, $\beta_1>1$). It follows from (\ref{eq:eq4}) that the ratio
of the variability periods in the optical and radio will be
\begin{equation}
	\frac{P_1}{P_2}=\frac{\delta \left( \theta_0, \beta_2 \right)}{\delta \left( \theta_0, \beta_1 \right)}.
	\label{eq:eq5}
\end{equation}
We used (\ref{eq:eq2}) and (\ref{eq:eq5}) to write the equation
\begin{equation}
	\frac{1-\beta_1 \cos \theta_0}{1-a \beta_1 \cos \theta_0} \sqrt{\frac{1-a^2 \beta_1^2}{1-\beta_1^2}}=\frac{P_1}{P_2},
	\label{eq:eq6}
\end{equation}
which was solved numerically for $a$ for values $\beta_1 =0.3-0.9999$ (corresponding to Lorentz factors $\Gamma =1-70$) with $P_1 = 3.3$~yrs and $P_2 = 5.8$~yrs.

We found that one root is always greater than one.
The other root is negative when $\beta_1<0.52$ and satisfies
our conditions when $\beta_1\geqslant 0.52$. When $a\approx 0.962$
and $\beta_1=0.9948$, the maximum value $\beta_2\approx0.957$ is
reached, corresponding to $\Gamma= 3.5$.
Solving (\ref{eq:eq6}) for $P_2$ and the variation period for the position angle of
the inner jet for this interval of $\beta_1$ values yields the
maximum value $\beta_2 = 0.949$ (for $a\approx 0.954$ and $\beta_1 =0.995$), which corresponds to $\Gamma = 3.1$. Thus, agreement
of the variability periods observed at difference
distances from the jet base can be achieved when $\Gamma\sim3-4$. 
This does not agree with the values $\Gamma\sim 10-20$
inferred from observations of superluminal motions of
jet components in 0716+714 \citep[see, e.g.,][]{Bach05,Nesci05,Pushkarev09}.

\subsection{Non-Radial Motion of the Jet Components}

\begin{figure}
    \centering
    \includegraphics[scale=0.8]{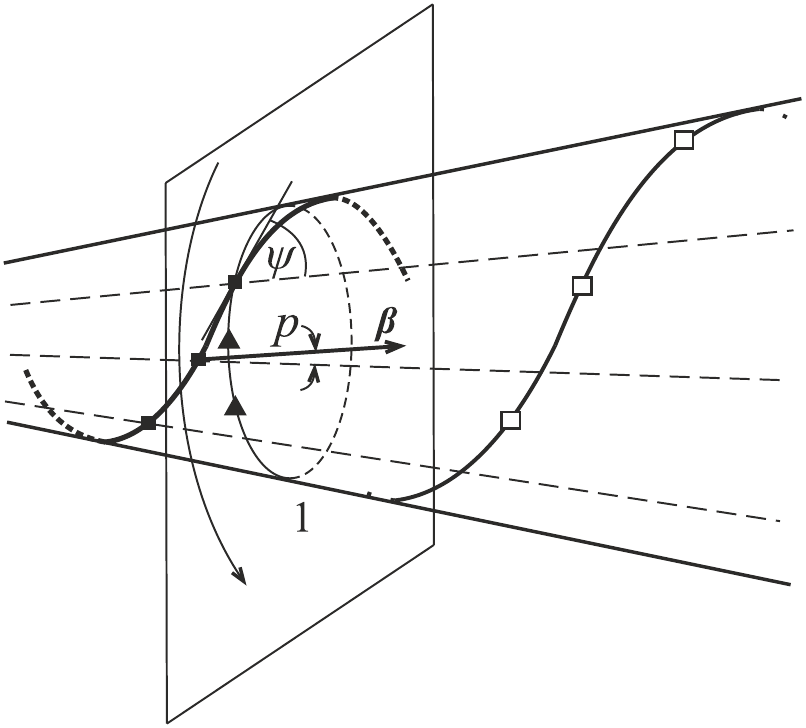}
    \caption{Schematic of a helical jet with non-ballistic motion of its components (shown by the filled squares). Part of the helical
path of the jet located near circle \textit{1} is shown (bold curve). The thin curves show the position of this part of the jet and the
components at some later time, with $\psi \neq p$. The triangles on circle~\textit{1} show the places where the jet components cross circle~\textit{1}.
The arrows show the direction of motion of a notional point along circle \textit{1}.}
    \label{fig:fig3}
\end{figure}

Let us now consider a helical jet whose components
move non-ballistically, i.e., at some angle to a
radial trajectory. We denote $p$ to be the pitch angle
(angle between the generating cone and the velocity
vector of a jet component), and $\psi$ to be the angle
between the tangent to the helix of the jet and the
generating cone at a given point (Fig.~\ref{fig:fig3}).
If $p=\psi$, the jet will appear stationary in space, and will always
cross circles \textit{1}, \textit{2}, and \textit{3} at the same points. In this
case, there should be no periodic variability, since the
angle $\theta$, and consequently the Doppler factor, do not
change in the regions responsible for the observed
quantities.
If $p\neq \psi$, we will observe the jet helix
rotating about its axis. Due to the conical geometry
of the jet, the variations of the azimuth angle $\varphi$ decrease
with increasing $r$.
However, we are interested
in variations of $\varphi$ at the constant distances $r_i$ from the
cone apex to the circles $i$ corresponding to the regions
making the main contributions to the optical ($i = 1$)
and radio ($i = 2$) emission, and to the region where
PA$_\text{in}$ can be measured ($i = 3$) (Fig.~\ref{fig:fig1}).

\begin{figure}
    \centering
    \includegraphics{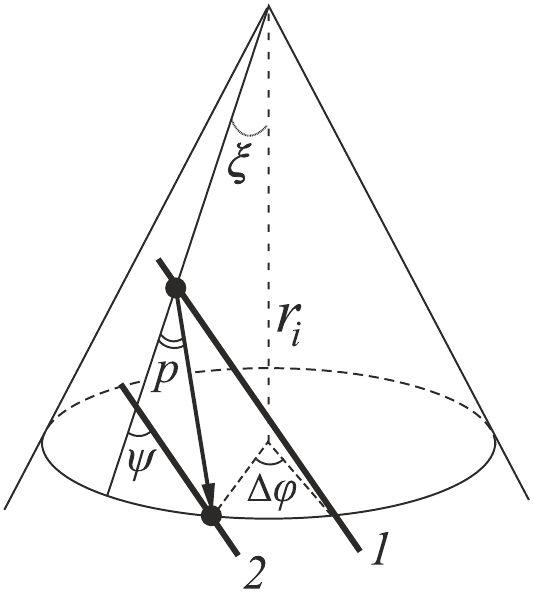}
    \caption{Schematic of part of the jet (bold curves) at a
distance $r_i$ from the cone apex at the initial time (\textit{1}) and
after a time dt (\textit{2}).}
    \label{fig:fig4}
\end{figure}

Variations of the azimuthal angle of the part of the
jet reaching a given distance $r_i$ from the cone apex
can be found from the schematic presented in Fig.~\ref{fig:fig4},
under the condition that $\beta c\, dt r_i$:
\begin{equation}
    d\varphi\approx\frac{\beta c \,dt\,\sin(\psi-p)}{r_i \sin \xi \cos \psi},
    \label{eq:eq7}
\end{equation}
where $\xi$ is the opening angle of the cone \citep[$\xi=1^\circ$,][]{Butuzova18}.
Since the angular frequency of a notional point moving along circle $i$ is $\omega_i = d \varphi / d t$, we find that the ratio of the periods of two observable quantities $i$ and $k$ is equal to the ratio of the distances from the cone apex to the region of the jet where these quantities are measured:
\begin{equation}
    \frac{P_i}{P_k}=\frac{r_i}{r_k}.
    \label{eq:ratioP}
\end{equation}

Substituting various pairs of the known variability periods for 0716+714 into (\ref{eq:ratioP}) yields the three independent relations
\begin{equation}
\begin{split}
    r_2 = & 1.76\, r_1,\\
    r_3=&3.30\, r_1, \\
    r_3=&1.88\, r_2.
\label{eq:express-r}
\end{split}
\end{equation}
It follows from the last two equations of (\ref{eq:express-r}) that $r_2 /r_1 \approx 1.76$, which is equal to the directly inferred ratio $r_2/r_1$ in the first equation of (\ref{eq:express-r}).
Thus, the observed periods for the long-term variability in the ratio, optical and inner-jet position angle PA$_\text{in}$ show good consistency. 
This supports a picture with non-radial motion of the jet features with $p \neq \psi$ and an absence of deceleration at the distances from the active nucleus considered here.

Let us suppose that PA$_\text{in}$ is measured at a specified distance from the VLBI core at 15~GHz, equal to 0.15~mas. 
Then, $r_3 - r_2 = 0.15$(mas)$/ \sin \theta_0$.
Using the third equation of (\ref{eq:express-r}) and $\theta_0=5.3^\circ$ \citep{Butuzova18}, we obtain the distance $r_2 = 1.84$~mas.
In a $\Lambda$CDM model with $H_0 = 71$~km~s$^{-1}$Mpc$^{-1}$, $\Omega_m=0.27$, and
$\Omega_\Lambda = 0.73$ \citep{Komatsu09} and adopting a redshift of $z = 0.3$ for 0716+714, the physical distance from the cone apex to the position of the VLBI core is 8.1~pc.
This is consistent with the distance of the VLBI core from
the black hole of 6.68~pc at 15.4~GHz determined by \citet{Pushkarev12} using these same cosmological parameters. 
This provides additional support for our picture of the jet.
Continuing our reasoning using (\ref{eq:express-r}), we find that the distance between the jet apex and the region where the optical emission becomes observable is 4.6~pc.
Due to the small delay in the variability \citep[at the limit of the time resolution of high frequency data of][]{Rani13, Larionov13}, we infer that the gamma-ray and optical emission is formed in the same region, or at least in closely spaced regions.
\citet{Rani14} estimated that the gamma-ray emission arises from a region located $3.8\pm1.9$~pc closer to the black hole relative to the VLBI core (observed at 43 and 86~GHz), also consistent with our results.

\section{Correlation between flux and inner-jet position angle}

Assuming that the long-term variability of the
blazar 0716+714 is due to periodic variations in the
direction of motion of the jet components, we expect
there should be a relationship between the measured
spectral flux density $F_\nu$ and the inner-jet position
angle. Such relationships between the gamma-ray
flux $F_\gamma$ and both the flux from the VlBI core at 43
and 86~GHz and PA$_\text{in}$ have been investigated for
0716+714 by \citet{Rani14}. They found that time
intervals with a strong positive correlation between
$F_\gamma$ and PA$_\text{in}$ alternate with intervals in which there
is a strong negative correlation between these two quantities.

This result was explained by \citet{Rani14} by the
fact that, in a curved (possibly helical) jet, the regions
responsible for the observed quantities are located at
different distances from the active nucleus, and therefore
have different viewing angles $\theta$. If the $\theta$ values
for two regions are roughly the same, there will be a
strong positive correlation between the corresponding
observed quantities. If the $\theta$ values are different, there
may be a strong negative correlation. However, for
the helical jet we are considering here, there is no
direct relationship between $F_\nu$ and PA$_\text{in}$, which can
be explained as follows. According to \citet{Butuzova18} (Eq. (1)),
the position angle is
\begin{equation}
\text{PA}_\text{in}=\text{PA}_0+\Delta \text{PA} = \text{PA}_0+\frac{\sin \xi \sin \varphi}{\cos \xi \sin \theta_0+\sin \xi \cos \theta_0 \cos \varphi}
    \label{eq:eq10}
\end{equation}
(PA$_0$ is the mean value of PA$_\text{in}$) and the angle between
the component velocity and the line of sight depend
on the geometrical parameters of the cone and the
position of the component relative to the cone axis and
the line of sight (i.e., on the angle $\varphi$).
We can use these equations, where the periodicity appears only due to variations in the azimuthal angle, to model the observed correlation between the flux and PA$_\text{in}$.

\begin{figure}
    \centering
    \includegraphics{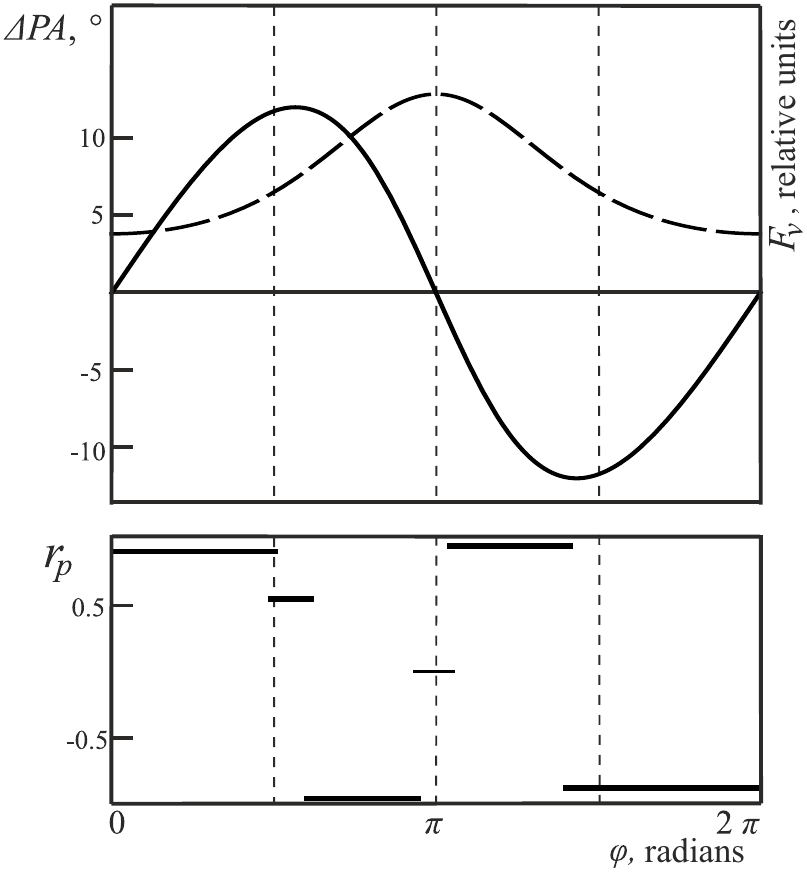}
    \caption{Deviations of the position angle of the inner jet
from its mean value $\Delta$PA (solid curve) and the observed
flux $F_\nu$ (dashed curve) over the variation period for $\varphi$
(upper), and the correlation coefficient $r_p$ between these
quantities (lower) for ballistic motion of features of a
helical jet.}
    \label{fig:fig5}
\end{figure}

Let us first consider the case of radialmotion of the
components. We find from (1) with the substitution
of (2), in which \citep[see][Eq. (5)]{Butuzova18}
\begin{equation*}
    \theta=\theta_b=\arccos \left( \cos \xi \cos \theta_0 - \sin \xi \sin \theta_0 \cos \varphi \right),
\end{equation*}
that the extrema of the function $F_\nu$ occur at values
$\varphi=n \pi$ (maxima for $n= 1$, 3, 5, etc. and minima for $n = 0$, 2, 4, etc.).
That is, the qualitative variations of $F_\nu$ do not depend on the choice of $\nu$, $\alpha$, $s$, and $\beta$.
The upper panel of Fig.~\ref{fig:fig5} presents the variations of $F_\nu$ calculated using (\ref{eq:eq1}) and (\ref{eq:eq10}) (for $\alpha=0.5$, $s = 2$, and $\beta = 0.995$, which corresponds to $\Gamma=10$)
and deviations of the inner-jet position angle from
its mean value $\Delta$PA (for $\theta_0/\xi=5.3$) as functions of the azimuthal angle $\varphi$.
The flux is normalized so as
to enable a visual comparison of its variations with
the variations of $\Delta$PA.
The resulting curves were  divided into several sections in $\varphi$, such that qualitative
variations in the behavior of both quantities did not
arise within each section.

The corresponding formulae were used to compose
datasets of $\Delta$PA and $F_\nu$ values for each section, for variations of $\varphi$ in steps of $0.5^\circ$, and the Pearson
correlation coefficient ($r_p$) between these datasets
was calculated. The resulting $r_p$ value will have its
maximum possible value, since, in contrast to the
observational data, there is no measurement error.
An alternation of intervals of strong positive and negative correlations can be seen (Fig.~\ref{fig:fig5}, lower panel).
Intermediate values of the correlation coefficient are
present only in short intervals. This theoretical result
is in qualitative agreement with the observations of
the behavior of the gamma-ray flux and the inner-jet
position angle considered by \citet{Rani14}.
Thus, we found that, for the same flux value (corresponding to the same viewing angle $\theta_b$), either a positive or negative correlation with PA$_\text{in}$ could be present in the observations.

\begin{figure}
    \centering
    \includegraphics{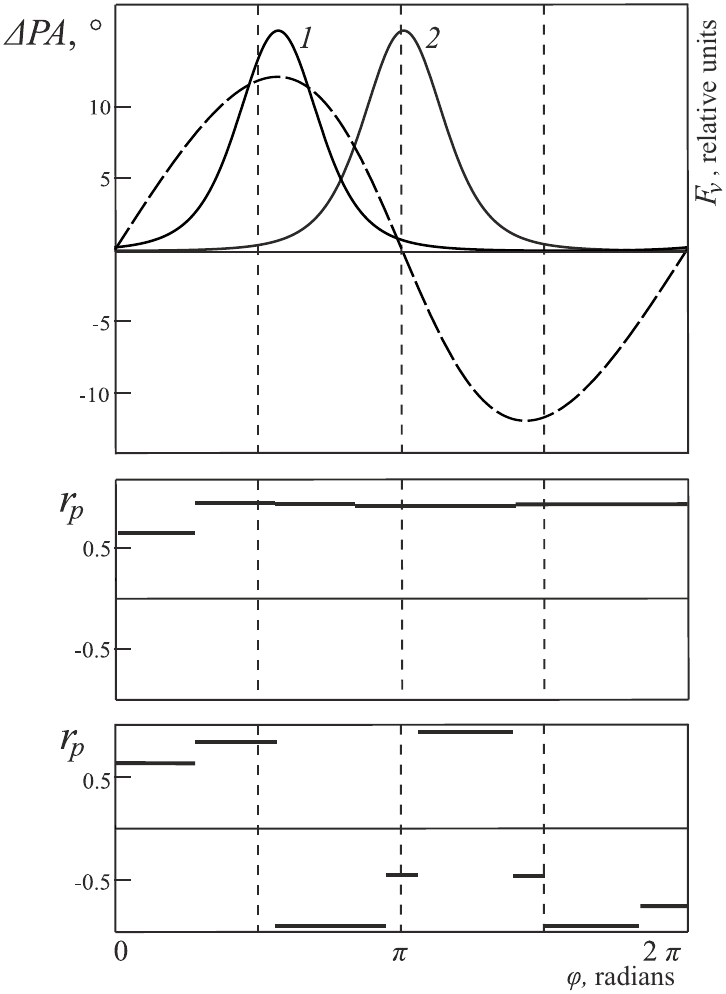}
    \caption{Upper: variations of the inner-jet position angle
relative to the mean value $\Delta$PA (dashed curve) and of the
flux $F_\nu$ for the case of zero difference in the azimuthal angles
of the regions responsible for the observed quantities
(curve~\textit{1}) and for $\Delta \varphi \approx 79^\circ$ (curve~\textit{2}). The central and
lower panels show the correlation coefficients $r_p$ for the
former and latter cases, respectively.}
    \label{fig:fig6}
\end{figure}

In the case of non-radial component motions,
the variations of the viewing angle are given by
Eqs. (11)-(13) from \citep{Butuzova18}, which we do not present
here due to their unwieldiness. The extrema of the
function $F_\nu$ occur at the values
\begin{equation}
    \varphi=-\arcsin \left( \frac{\sin p}{\sqrt{\cos^2 p \sin^2 \xi+\sin^2 p}} \right)+\pi n,
    \label{eq:eq11}
\end{equation}
where even $n$ correspond to maxima and odd $n$ to
minima of the function $F_\nu$. According to (\ref{eq:eq11}), for the pitch angle found by \citet{Butuzova18} of $p = 5.5^\circ$, the flux reaches a maximum when $\varphi \approx 100^\circ$.
As $p$ is decreased, the
peak $F_\nu$ shifts toward 180$^\circ$, and the maximum $F_\nu$
occurs for $\varphi \approx 90^\circ$ when $p > 20^\circ$.
The upper panel of Fig.~\ref{fig:fig6} plots the functions $F_\nu$ and $\Delta$PA for the same parameters as in the previous case. The correlation coefficient $r_p$ was constructed using an analogous
procedure (Fig.~\ref{fig:fig6}, middle panel).
This figure shows
that a strong positive correlation between the inner jet
position angle and the observed flux should always
be present. However, since these quantities arise in
regions located at different distances from the jet base
in our model (see Section~2), we can conclude with
confidence that the azimuthal angles of the components
for which $F_\nu$ and $\Delta$PA are observed at a given
time differ by some amount $\Delta \varphi$.

Agreement with the results of \citet{Rani14}
requires that this difference be $\Delta \varphi \approx 79^\circ$ (curve~\textit{2} for $F_\nu$ in the upper panel and curve for $r_p$ in the lower panel in Fig.~\ref{fig:fig6}).
On the other hand, strong
positive or negative correlations between $F_\nu$ and $\Delta$PA
are also possible for radial component motions if the
difference in the azimuthal angles is $\Delta \varphi \approx \pi/2$ or $\approx \pi$ (see Fig.~\ref{fig:fig5}).
Consequently, the correlation between
the observed quantities may be insignificant when
analyzing data over long time intervals, as was found
by \citet{Raiteri03}, for example, in their analysis of
the radio and optical fluxes during 1994$-$2001.
In contrast, the data for the shorter interval 2007$-$2010
analyzed by \citet{Rani13} revealed a correlation between
the indicated quantities at a significance level
of more than 99$\%$. Further, the difference in the
azimuthal angles of the regions responsible for the
observed quantities can appreciably affect both the
correlation coefficient between PA$_\text{in}$in and $F_\nu$ and the
duration of the time interval when a given correlation
coefficient is observed. Finally, the character of the
motions of individual components of helical jet cannot
be determined by analyzing the correlation between
PA$_\text{in}$ and $F_\nu$. It is important to note that, when investigating correlations between fluxes observed in different spectral ranges, distinct correlation coefficients
in the different time intervals will also be present.

\section{Discussion and conclusion}

The hypothesis that AGN jets may be helical has
been widely applied for several decades to interpret
various observed properties such as their microvariability
\citep{CamKrock92}, the shape and variations of the spectral
energy distributions for the blazars Mrk~501 \citep{VilRait99} and
0716+714 \citet{OstoreroRait01}, the long-term brightness variability of OJ~287 \citep{Sillanpaa88} and 0716+714 \citep{Nesci05}, variations in the
speeds and non-radial component motions \citep{Rastorgueva09}, and
quasi-periodicity of variations of the inner-jet position
angle for 0716+714 \citep{Bach05, Lister13}.
A helical jet shape could form due to precession of the jet nozzle or
the development of (magneto)hydrodynamical instabilities.
Kelvin–Helmholtz instability \citep{Hardee82} has been
widely studied as a means of estimating the physical
parameters of jets and the ambient medium (e.g., for
3C~120 \citep{Hardee03} and 0836+710 \citep{Perucho12}). Alternatively, wavelike perturbations at the boundaries of the observed isophotes of jets could be related to the development
of a magnetohydrodynamical analog of wind instability \citep{GestrinKont86}.

In this study, we have supposed that the radiating
jet components form a helical curve, without
considering their physical nature. For example, the
jet components could be individual radiating parts of
the jet (plasmoids or regions of shocks) or volume
elements (in the case of a spatially continuous radiating
jet). A helical shape suggests the presence of
periodic variations of the angle between the jet velocity
and the line of sight at some constant distance
from the core, which should be manifest as long-term
quasi-periodic variability of the radiation flux
over the entire observed range of the electromagnetic
spectrum of the blazar 0716+714. The differences
in the quasi-periods for the variations of PA$_\text{in}$ \citep{Bach05, Lister13} and of the radio- \citep{Raiteri03, Bychkova15, Bach05,LiuMi12} and optical \citep{Raiteri03}
fluxes can most simply be explained in the jet geometry
considered if the radiation in different spectral
ranges is emitted at different distances from the jet
apex. This spatial separation of regions radiating
at different frequencies can arise due to synchrotron
self-absorption in the jet or the energy losses of the
radiating electrons. In both cases, the higher the
frequency of the observed emission, the closer to the
jet base the region in which it is generated. The delays
in the flux variability observed at different frequencies
also testify to the action of this effect \citep[see, e.g.,][]{Raiteri03,Larionov13,Rani14}.

In this study, we have brought the variability periods
in different spectral ranges into agreement in the
case of radial and non-radial motions of the radiating
components. The former case requires a low Lorentz 
factor for the components (no more than 4) with
overall deceleration, at least from the region where
the optical emission is formed to 0.15-0.5 mas from
the core, where the position angle of the inner jet is
measured. This does not agree with observations of
features in the parsec-scale jet of 0716+714 \citep{Bach05,Nesci05, Pushkarev09}.
Moreover, it was shown by \citet{Butuzova18} that differences between the relative speeds of components
in the inner and outer jet observed in different years
can be explained in the framework of a helical-jet
model with non-ballistic component motions, such as
are observed in the jet \citep{Bach05,Rastorgueva09}.
It was shown that period ratio is equal to ratio of the physical distances from the jet apex of the regions responsible for the measured quantities. This enables us to introduce another absolute distance scale, which will subsequently facilitate deeper studies of the jet properties.
In our picture of the jet and the appearance of long-term variability due to geometric effects, the radio periods found by \citet{Bychkova15} cannot carry information about the properties of the central engine without taking into account the non-radial motions of the jet components.

The helical shape of a jet with spatially separated
regions responsible for the observed emission
at different frequencies and region where the inner jet
position angle is measured complicates studies of
correlations between the observed quantities. It has
been shown that there cannot be a constant correlation
coefficient between quantities formed in regions
at different fixed distances from the jet apex.
A strong positive correlation observed in one time interval will
be replaced with a negative correlation in another.
This is due to both the different azimuthal angles of
these regions and the fact that $\varphi$ varies irregularly in the observer's rest frame, due to variations of $\delta$, especially
for non-radial component motions \citep{Butuzova18}. This agrees with certain observational facts.

For example, \citet{Rani13} noted an alternation
of intervals when positive and negative correlations
between PA$_\text{in}$ and the gamma-ray flux of the blazar 0716+714 were observed. \citet{Rani13} also indicate that a strong correlation between the gamma-ray
and optical fluxes was observed over roughly
500 days (Pearson correlation coefficient $r_p = 0.66$), while $r_p = 0.36$ over the following $\approx$400 days.
A correlation was also found between the radio flux and
PA$_\text{in}$ during 1994$-$2014 ($r_p = 0.44$) \citep{LiuMi12}, while no correlation between these quantities was found for data obtained from August 2008 through September 2013 \citep{Rani14}.

We can introduce some clarity into this picture
only if we know the geometrical parameters of the
helical jet, the character of the motion of the jet components,and the arrangement of the studied regions
responsible for various observed quantities relative to
the plane containing the axis of the helical jet and
the line of sight, and not only relative to the line of
sight, as was supposed in the simplest case \citep{Rani14}. We also showed in Section~3 that it is not possible to
determine the character of the component motions in
a helical jet based on the correlation between the radio flux and the inner-jet position angle, as was done by \citet{LiuMi12}.

Our hypothesis that the jet of the blazar 0716+714
has the form of a helical curve located on the surface of a cone may seem somewhat idealized. However, this
is consistent with the results of many years of VLBI
observations \citep{Lister13,Pushkarev17}. In addition, a helical jet with non-radial component motions makes it possible to find agreement between estimates of the component velocities in the VLBI jet and the viewing angle obtained in different studies, which often differ appreciably \citep{Butuzova18}. As we have shown, such a jet can provide a
simple explanation for the differences in the observed
long-term quasi-periods for different quantities, as
well as the differences in the correlations between the
observed quantities in different time intervals.

\acknowledgments{The author thanks A.B. Pushkarev for useful comments.}

\bibliography{helicaljet0716}{}
\bibliographystyle{aasjournal}

\end{document}